\documentclass[traditabstract,letter]{aa}

\usepackage{graphicx}
\usepackage{txfonts}

\usepackage{natbib}
\bibpunct{(}{)}{;}{a}{}{,}

\usepackage{hyperref}
\hypersetup{colorlinks=true,
            citecolor=blue,
            linkcolor=blue}
            
\usepackage{tikz}

\def \be {\begin{equation}}
\def \ee {\end{equation}}
\def\vv#1{\vec{#1}}
\def\ep {\mathrm{e}}
\def\ii {\mathrm{i}}
\def\ia{a}
\def\jb{b}
\def\kc{c}
\def\m{m}
\def\w{\omega}
\def\vr{\vv{r}}

\def\vi{\vv{\hat \ia}}
\def\vj{\vv{\hat \jb}}
\def\vk{\vv{\hat \kc}}
\def\vw{\vv{\w}}
\def\vL{\vv{L}}
\def\vrb{\vv{r}_b}
\def\vrp{\vv{r}_0}
\def\vrc{\vv{r}_1}
\def\rb{r_b}
\def\rp{r_0}
\def\rc{r_1}
\def\b{b}
\def\ab{a_\b}
\def\nb{n_\b}
\def\eb{e_\b}
\def\fb{f_\b}
\def\Mb{M_\b}
\def\ve{\varepsilon}
\def\vo{\rho}
\def\vp{\varphi}

\def\bfx#1{#1}
\def\figpath{}
\def \llabel#1{\label{#1}}
\usepackage{color}

\begin{document}

\title{Spin-orbit coupling and chaotic rotation for circumbinary bodies\thanks{Fig. \ref{fig:tides} and Appendices \ref{eom} and \ref{swe} are available in electronic form at \url{http://www.aanda.org}}} % in quasi-circular orbits.}
\subtitle{Application to the small satellites of the Pluto-Charon system}
%\subtitle{none}
\titlerunning{Spin-orbit coupling for circumbinary bodies}

\author{
Alexandre C. M. Correia\inst{1,2}
\and Adrien Leleu\inst{2}
\and Nicolas Rambaux\inst{3,2}
\and Philippe Robutel\inst{2}
}

%\offprints{}
 
\institute{
CIDMA, Departamento de F\'isica, Universidade de Aveiro, Campus de
Santiago, 3810-193 Aveiro, Portugal
  \and 
ASD, IMCCE-CNRS UMR8028,
Observatoire de Paris,
77 Av. Denfert-Rochereau, 75014 Paris, France  
  \and 
Universit\'e Pierre et Marie Curie, UPMC - Paris 06 
}

\date{Received ; accepted To be inserted later}
% The correct dates will be entered by the editor

\abstract{
We investigate the resonant rotation of circumbinary bodies in planar quasi-circular orbits.
Denoting $\nb$ and $n$ the orbital mean motion of the inner binary and of the circumbinary body, respectively, we show that spin-orbit resonances exist at the frequencies $n\pm k\nu/2$, where $\nu = \nb - n$, and $k$ is an integer.
Moreover, when the libration at natural frequency has the same magnitude as $\nu$, the resonances overlap and the rotation becomes chaotic.
We apply these results to the small satellites in the Pluto-Charon system, \bfx{and conclude that their rotations are likely chaotic. 
However, the rotation can also be stable and not synchronous for small axial asymmetries.}
}


   \maketitle
%
%________________________________________________________________

% + frequente no Google:
% high/low inclination (alta/baixa inclinacao)
% high/low eccentricity (alta/baixa excentricidade)

% alternativa:
% large/small eccentricity (grande/pequena excentricidade)

%%%%%%%%%%%%%%%%%%%%%%%%%%
\section{Introduction}

%At present, several circumbinary bodies are known. 
Circumbinary bodies are objects that orbit around a more massive binary system.
In the solar system, the small satellites of the Pluto-Charon system are the best example \citep[e.g.,][]{Brozovic_etal_2015}.
Planets orbiting two stars, often called circumbinary planets, have also been reported \citep[e.g.,][]{Correia_etal_2005, Doyle_etal_2011, Welsh_etal_2012}.
Most of these bodies are close enough to the central binary to undergo tidal dissipation,
which slowly modifies the rotation rate until it becomes close to the mean motion \citep[e.g.,][]{MacDonald_1964, Correia_etal_2014}.

%aqui
%When the rotation rate and the mean motion have the same magnitude, %($\dot\theta\sim n$), the dissipative tidal torque may be counterbalanced by the conservative torque due to the axial asymmetry of the inertia ellipsoid.
%aqui
For an object that moves in an eccentric orbit around a single massive body, the rotation rate can be captured in a half-integer commensurability with the mean motion, usually called a spin-orbit resonance \citep{Colombo_1965, Goldreich_Peale_1966}. %: $\dot \theta = n/2, \, n, \, 3n/2, \, 2n$, and so on.
Moreover, for very eccentric orbits or large axial asymmetries, 
the rotational libration width of the individual resonances may overlap, and 
the rotation becomes chaotic \citep{Wisdom_etal_1984,Wisdom_1987}.
However, for nearly circular orbits, all these equilibria disappear % \citep[e.g.,][]{Correia_Laskar_2009}
and the only possibility for the spin is  synchronous %spin-orbit resonance
rotation \citep[e.g.,][]{Correia_Laskar_2009}.%  \citep[e.g.,][]{Goldreich_Peale_1966}. %, Wisdom_etal_1984,Wisdom_1987}. %: $\dot\theta=n$.
%Since tidal dissipation simultaneously damps the eccentricity to zero \citep[e.g.,][]{Hut_1980, Correia_2009}, all the main satellites in the solar system are observed in nearly circular orbits and synchronous rotation. %, including the Moon.
%Saturn coorbital satellites also present very low eccentricities (less than 0.01) and their rotations appear to be synchronous, although there is yet no confirmation \citep{Tiscareno_etal_2009}.

When a third body is added to the classic two-body problem, the mutual gravitational perturbations introduce additional spin-orbit resonances at the perturbing frequency \citep{Goldreich_Peale_1967, Correia_Robutel_2013}.
In the case of circumbinary bodies there is, in addition, a permanent misalignment of the long inertia axis of the rotating body from the radius vector pointing to each inner body (Fig.\,\ref{ref:angles}).
The resulting torque on the rotating body's figure induces some rotational libration that may give rise to some unexpected behaviors for the rotation rate \citep{Showalter_Hamilton_2015}.
In this Letter we investigate all possibilities for the rotation of circumbinary bodies, and apply them to the spin evolution of the small satellites of the Pluto-Charon system.

\vskip-0.05cm
%%%%%%%%%%%%%%%%%%%%%%%%%%
\section{Spin-orbit coupling}
\llabel{sor}

We consider a planar three-body hierarchical system composed of a central binary with masses $m_0$ and $m_1$, together with an external circumbinary companion with mass $m$ (Fig.\,\ref{ref:angles}), where $\m \ll m_1 \le m_0 $. %$\m \ll (m_0+m_1)$ 
For the orbits we use Jacobi canonical coordinates, with $\vrb $ being the position of $m_1$ relative to $m_0$ (inner orbit), and $ \vr $ the position of $\m$ relative to the center of mass of $m_0$ and $m_1$ (outer orbit).
The body with mass $\m$ has principal moments of inertia $ A \le B \le C $ and angular velocity given by $\vw$.
\bfx{In Appendix~\ref{eom} we provide the full equations of motion for the spin.
Since tidal dissipation usually damps the obliquity to zero \citep[e.g.,][]{Hut_1980, Correia_2009}, for simplicity, here we describe the motion for $\vw$ normal to the orbit.}
In absence of tides, the equation for the rotation angle, $\theta$, is then given
by (Eq.\,(\ref{150626b}))
\be
\ddot \theta = -\frac{\sigma^2}{2} \left[ (1-\delta) \left(\frac{a}{\rp}\right)^3 \sin 2 (\theta - f_0) + \delta \left(\frac{a}{\rc}\right)^3 \sin 2 (\theta-f_1) \right] \ ,
\llabel{150616a}
\ee
where $a$ and $n$ are the semi-major axis and the mean motion of the outer orbit, respectively, 
($r_i, f_i$) are the radial and the angular coordinates of $\vr_i$,
\be 
\vrp = \vr + \delta \vrb  \quad  \mathrm{and} \quad \vrc = \vr + (\delta-1) \vrb \ ,
\llabel{150616b}
\ee
\be
\delta = \frac{m_1}{m_0+m_1} \ , \quad \mathrm{and} \quad
\sigma = n\sqrt{3\frac{B-A}{C}} \ .
\llabel{150616c}
\ee
The parameter $\delta$ is the mass ratio of the inner binary, %($\delta = 0$ for $m_1=0$, and $\delta = 1$ for $m_0=0$), 
while $\sigma$ is approximately the frequency of small-amplitude rotational librations in an unperturbed synchronous resonance.

\begin{figure}[ht]
\begin{center}
\includegraphics[width=.8\columnwidth]{\figpath 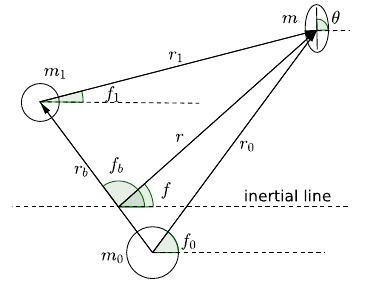}
 \caption{Reference angles and coordinates. The system is coplanar and is composed of an inner binary with masses $m_0$ and $m_1$, and an outer body $\m$ that rotates \bfx{with angular velocity $\vw$}.
 %about an axis normal to the orbits.  
 \llabel{ref:angles} }
\end{center}
\end{figure}

We can express $\rp^{-3}$, $\sin f_0$, and $\cos f_0$ as
\be 
\rp^{-3} = ||\vr + \delta \vrb||^{-3} = r^{-3} \left(  1 + 2\delta \frac{\vr \cdot \vrb}{||\vr||^2} + \delta^2  \frac{||\vrb||^2}{||\vr||^2}  \right)^{-3/2} \ ,
\llabel{150617z}
\ee
and
\be 
\ep^{\ii f_0} = \ep^{\ii f} \left( 1 + \delta \frac{||\vrb||}{||\vr||} \ep^{\ii (\fb-f)} \right) \left(  1 + 2\delta \frac{\vr \cdot \vrb}{||\vr||^2} + \delta^2  \frac{||\vrb||^2}{||\vr||^2}  \right)^{-1/2} \ .
\llabel{150617y}
\ee
Similar expressions can be found for $\rc^{-3}$, $\sin f_1$, and $\cos f_1$, if we replace $\delta$ by $(\delta-1)$.
Therefore, since $||\vrb|| < ||\vr||$, we can develop equation (\ref{150616a}) in power series of $\rb/r$.
In addition, for the hierarchical problem the orbits are approximately ellipses.
Thus, $r$ and $\rb$ can also be expanded in terms of Hansen coefficients \citep[e.g.,][]{Laskar_Boue_2010}. 
We then rewrite equation (\ref{150616a}) as
\be
\ddot\theta = -\frac{\sigma^2}{2} \sum_{l,p,q} \beta_{l,p,q} (e, \eb, \delta, \alpha) \sin2 (\theta - p M - q \Mb) \ ,
\label{eq:rot_gen}
\ee
where $e$ is the eccentricity; $M$ is the mean anomaly; $\beta_{l,p,q} \propto \alpha^l $, with  $\alpha = \ab / a$; and $p$, $q$ are half-integers.
%Spin-orbit coupling is then possible whenever $\dot\theta = p n + q \nb$. 
Spin-orbit resonances occur whenever $\dot\theta = p n + q \nb$. 
In Appendix~\ref{swe} we provide the complete expression for equation (\ref{eq:rot_gen}) to the first order in the eccentricities and to the second order in $\alpha$.
%We see that series over terms in a linear combination of the angles $\theta$, $M$ and $\Mb$

%%%%%%%%%%%%%%%%%%%%%%%%%%
\section{Near circular orbits}
\llabel{circorb}

For near circular orbits ($e \approx 0$ and $\eb \approx 0$), we can neglect both eccentricities, thus $r \approx a$ and $\rb \approx \ab$.
Retaining terms in $\alpha^3$ in equation (\ref{eq:rot_gen}) gives

\begin{eqnarray}
\ddot \gamma  &=&   -\frac{\sigma^2}{2} \left[ ( 1 + \frac54\vo_2  ) \sin 2\gamma \right.\nonumber \\
&& + \frac38\vo_2\sin (2\gamma - 2\phi ) +\frac{35}{8}\vo_2\sin (2\gamma +2\phi) \nonumber \\
&& + \frac{15}{16}\vo_3\sin (2\gamma -\phi) +  \frac{35}{16}\vo_3\sin (2\gamma +\phi) \nonumber \\
&& \left. + \frac{5}{16}\vo_3\sin (2\gamma -3\phi ) +  \frac{105}{16}\vo_3\sin (2\gamma +3\phi) \right] 
%\nonumber \\ && + \, {\cal O}(e,\eb,\alpha^4)
\ , \llabel{150616z}
\end{eqnarray}
where $\gamma = \theta - M$, $\phi = \Mb - M$,
\be
\vo_2 = \delta(1-\delta) \, \alpha^2 \ , \quad \mathrm{and} \quad
\vo_3 = \delta(1-\delta)(1-2\delta)\, \alpha^3 \ .
\llabel{150616y}
\ee

We see that we have several islands of rotational libration,  
\be
\dot\gamma=\dot\theta-n=0,\pm \nu, \pm\nu/2, \pm 3\nu/2 \ , \quad \mathrm{with} \quad \nu = \dot \phi = \nb - n \ .
\llabel{150616x}
\ee
Each term individually behaves like a pendulum where the rotation can be trapped.
Therefore, together with the classic synchronous equilibrium at $\dot\theta=n$, there are additional possibilities for the spin at the super- and sub-synchronous resonances $\dot\theta=n + k\nu/2$, with $k\in\mathbb Z$.

Terms with $k\ne0$ depend on $\sigma^2$, $\delta$, and $\alpha^2$, that is, on the axial asymmetry of the body, on the mass ratio of the central binary, and on the semi-major axis ratio.
Thus, for circumbinary bodies far from the inner binary ($\alpha \ll 1$), the amplitude of these higher order terms decreases very quickly to zero.
For $\delta \approx 0$ (i.e., $m_1 \ll m_0$) we have $\vo_2 = \vo_3 = 0$, thus only the synchronous resonance subsists, as in the classic circular two-body problem \citep[e.g.,][]{Goldreich_Peale_1966}.
%The same occurs for $\delta \approx 1$ (i.e., for $m_0 \ll m_1$).
The effect of the binary mass ratio is maximized for $\delta = 1/2$ (i.e., $m_0 = m_1$), but only for the second order resonances (terms in $\vo_2$), since third order resonances also vanish for equal masses ($\vo_3 = 0$).
Interestingly, the size of the sub-synchronous resonance with $k<0$ is always larger than the symmetrical super-synchronous resonance with $k>0$.

%%%%%%%%%%%%%%%%%%%%%%%%%%
\subsection{Chaotic motion}
\llabel{chaos}

When the libration amplitudes of some individual resonant islands overlap, the rotational motion can be chaotic \citep{Chirikov_1979, Morbidelli_2002}.
This means that the rotation exhibits random variations in short periods of time.
This phenomenon was first described for the rotation of Hyperion, which is chaotic because its orbit is eccentric \citep{Wisdom_etal_1984}.

In the circular approximation, each individual resonance is placed at $\dot\gamma = k \nu/2$, with libration width $\sigma \sqrt{\beta_k}$, where $\beta_k = \sum_l \beta_{l,(1-k)/2,k/2}$.
Overlap between two resonances with $k_1 < k_2 $ then occurs whenever 
\be
k_1 \nu / 2 + \sigma \sqrt{\beta_{k_1}} \gtrsim k_2 \nu / 2 - \sigma \sqrt{\beta_{k_2}} 
%\Leftrightarrow \sqrt{\beta_{k_1}} + \sqrt{\beta_{k_2}} > (k_2-k_1) \frac{\nu}{2\sigma}
\ . \llabel{150618a}
\ee

The synchronous resonance ($k=0$) has the largest width, so overlap is more likely for this resonance.
The third order resonance with $k=-1$ is the nearest resonance with larger amplitude, thus chaos sets in for
\be
\sigma  \left( \sqrt{\frac{35}{16} \vo_3} + \sqrt{1+\frac54 \vo_2}  \right) \approx \sigma  \gtrsim \frac{\nu}{2} = \frac12 \left( \nb - n \right)
\ . \llabel{150618c}
\ee
%However, third order resonances are small, so chaotic motion is restricted to the separatrix of the synchronous resonance.
%Strong chaotic motion can only occur when the second order resonance $k=-2$ overlaps, that is, for
%\be
%\frac{\sigma}{n}  \left( \sqrt{\frac{35}{8} \vo_2} + \sqrt{1+\frac54 \vo_2}  \right) \approx \frac{\sigma}{n}  \gtrsim \frac{\nb}{n} - 1
%\ . \llabel{150618b}
%\ee

\bfx{Rotation in the chaotic zone is also attitude unstable \citep{Wisdom_etal_1984, Wisdom_1987}.
A non-zero obliquity introduces additional resonant terms to the rotation (Eq.\,(\ref{150616z})), increasing the chances of overlapping \citep[e.g.,][]{Correia_Laskar_2010}.
Therefore, in order to correctly account for the chaotic behavior, we need to integrate the full equations of motion (Appendix~\ref{eom}). }

%%%%%%%%%%%%%%%%%%%%%%%%%%
\subsection{Global dynamics}
\llabel{glodyn}

\begin{figure}
%%%%%%%%%%%%%%%%%%%%%%%%%%%%%%%%%%%%%%%%%%%%%%%%%%%%%%%%%%%%%%%%%%
%/Volumes/USER/robutel/Phil/articles-exposes/2013_Alexandre/calculs/fig_ampl_T_1_mu1em3_d1.t   retravaillŽe sous ilustrator.
% le fichier .dat est construit par fig_frq.t   avec fic_freq0 = "ampl_T_1_mu1em3_d1";  fic_ampl = "ampl_T.dat";
%%%%%%%%%%%%%%%%%%%%%%%%%%%%%%%%%%%%%%%%%%%%%%%%%%%%%%%%%%%%%%%%%%
\begin{center}
\includegraphics[width=1.\columnwidth]{\figpath 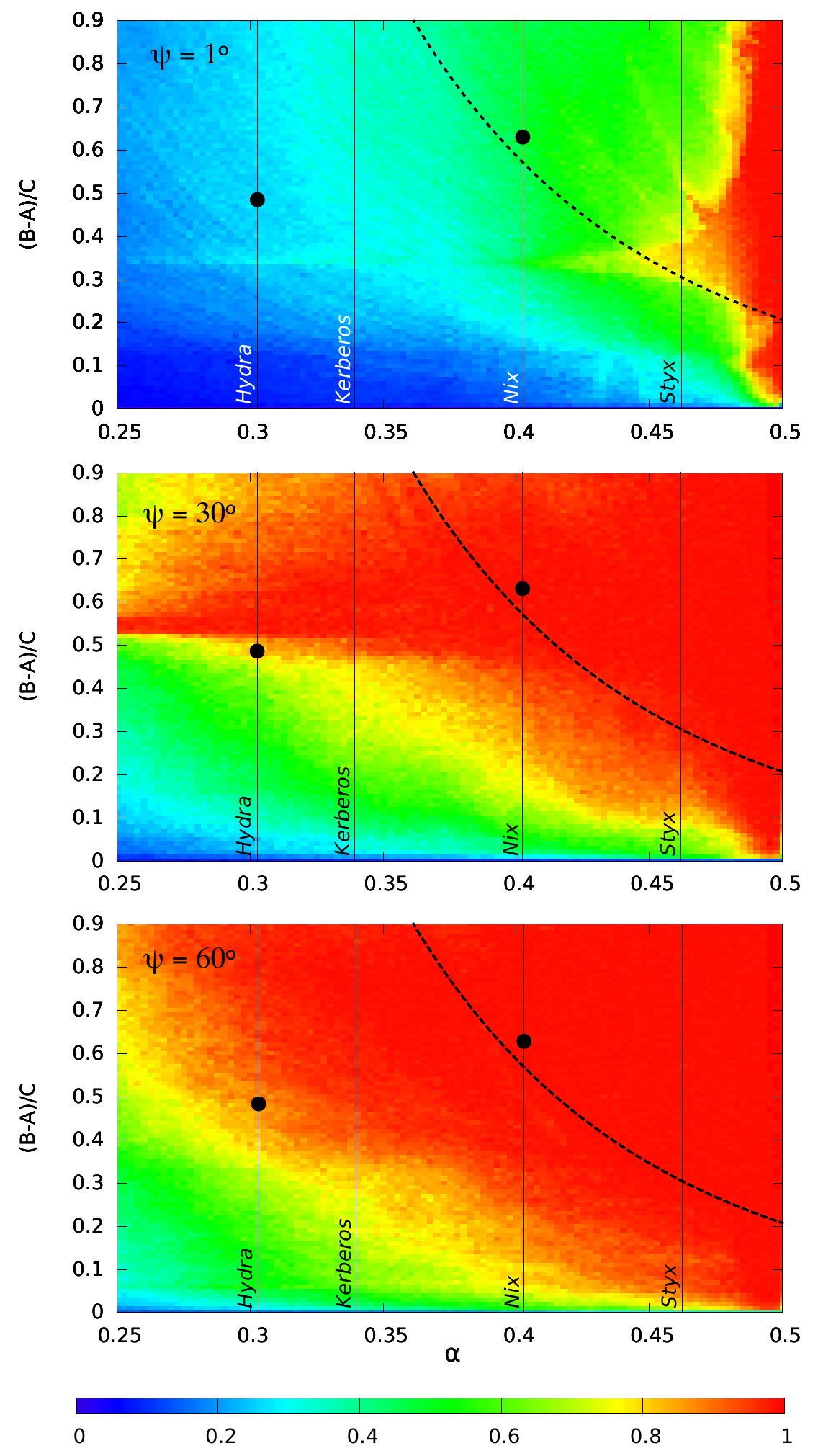} 
 \caption{Stability analysis of the rotation rate
close to the synchronization for $\delta = 0.1$, $\alpha \in [0.25:0.5]$, and $(B-A)/C \in [0:0.9] $. 
We adopt three different \bfx{initial obliquity values $\psi = 1^\circ$, $30^\circ$, and $60^\circ$.}
The color index indicates the proportion of chaotic orbits inside the studied domain: from dark blue for fully regular to red for entirely chaotic. 
The dashed line (Eq.\,(\ref{150618c})) roughly delimits the zone where chaotic motion is possible.
The dots identify two representative cases that we detail in Figure~\ref{fig:section}.
\label{fig:global}  }
\end{center}
\end{figure}

The general problem of the spin-orbit coupling for near circular circumbinary bodies can be reduced to the analysis of $\dot\gamma/n$, which depends on the libration frequency $\sigma/n$, the mass ratio $\delta$, and the semi-major axis ratio $ \alpha$ (Eq.\,(\ref{150616z})).
However, since $\delta \in [0,1/2]$ and it is a well-known parameter for a specific binary system, the global dynamics is approximately controlled by the only two free parameters: $\alpha$ (or $\nb/n$) and $ \sigma/n$ (or $(B-A)/C$). 

Rescaling the time with $n=1$, we can perform a stability analysis of $\dot\gamma$ in the plane ($ \alpha, \sigma$) to quickly identify the rotational regime for any system of quasi-circular circumbinay bodies that is near the synchronous equilibrium.
If isolated, the half-width of the synchronous resonant island in the direction of $\dot\gamma$ is equal to $\sigma$.
\bfx{When perturbed, chaos is expected for $\sigma \sim \nu$ (Eq.\,(\ref{150618c})).}
Thus, for a given ($ \alpha, \sigma$) pair, we fix the initial value of $\gamma$ at the libration center of the synchronous resonance, and select $400$ equispaced initial values of $\dot\gamma \in [-\nu:\nu]$. 
The corresponding solutions are integrated using the equations for the three-body problem together with \bfx{equations (\ref{150626a}) and (\ref{050104b}) for the rotational motion (without tidal dissipation).}
The dynamical nature (stable/unstable) is deduced from frequency analysis \citep{Laskar_1990,Laskar_1993PD}, which gives the fraction of chaotic trajectories. 

In Figure~\ref{fig:global} we show the general results \bfx{for $\delta=0.1$ using three different initial obliquity values $\psi = 1^\circ$, $30^\circ$, and $60^\circ$. 
All the remaining variables are initially set to zero.}
The color index indicates the proportion of chaotic orbits inside the studied domain: from dark blue for fully regular to red for entirely chaotic. 
Depending on the value of $\alpha$ and $\sigma$, the rotation can present different behaviors, ranging from non-synchronous equilibria to chaotic motion.
\bfx{Figure~\ref{fig:global} remains almost unchanged for $\delta > 0.1$} since the global dynamics is not very sensitive to $\delta$.
The only exception would be for $\delta \ll 0.1$ since in that case the system would behave like the classic circular two-body problem where the rotation can only be synchronous.

\begin{figure*}[ht]
\begin{center}
\includegraphics[width=1.\textwidth]{\figpath 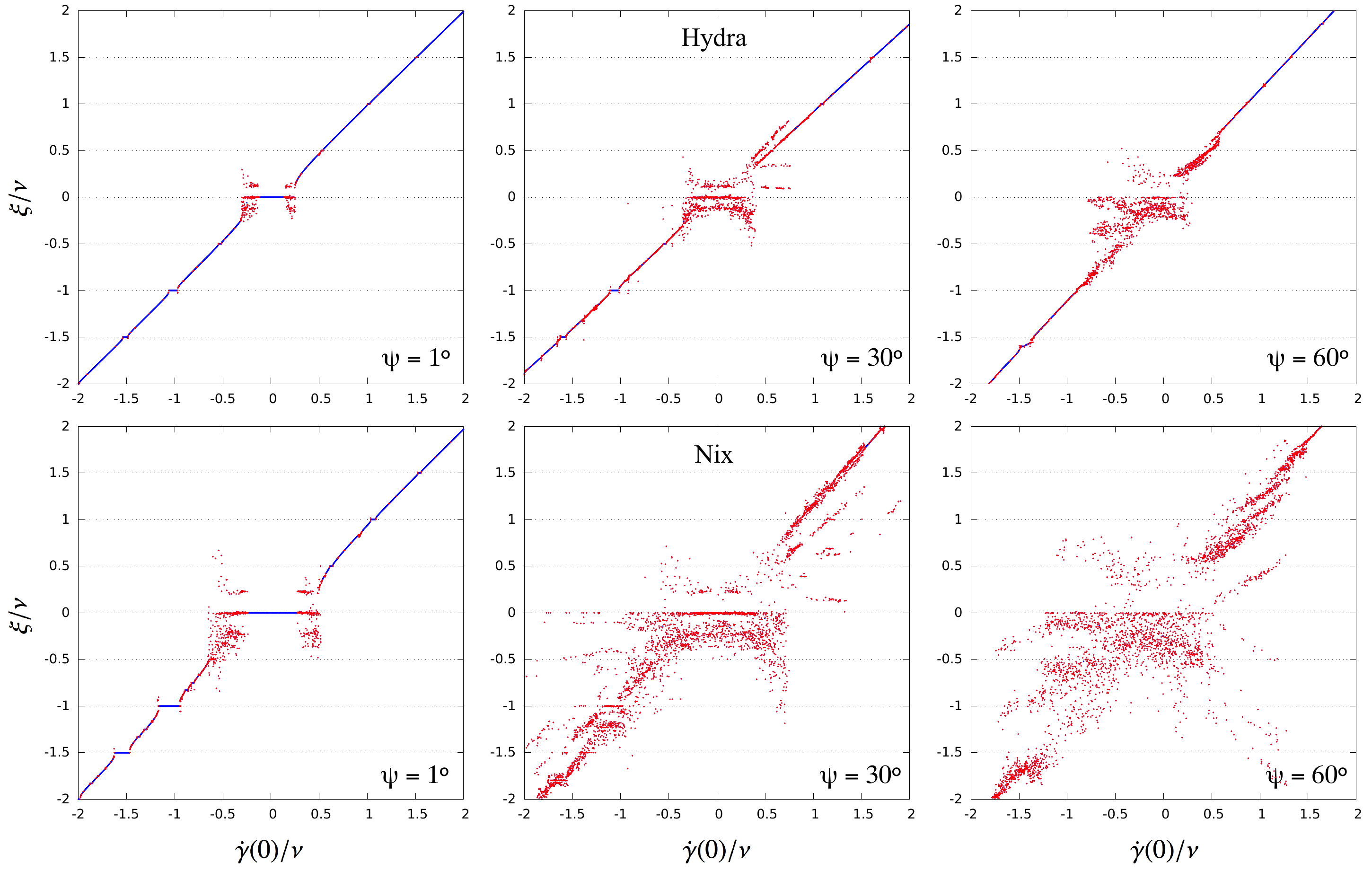}
  \caption{Cross sections representing the spin dynamics of \bfx{Hydra (top) and Nix (bottom) at different initial obliquities $\psi = 1^\circ$, $30^\circ$, and $60^\circ$}. The dots show the main frequency of $\gamma(t)$, denoted by $\xi$, for different initial values of $\dot \gamma(0)$. It is obtained by numerical integration of the three-body problem together with \bfx{equations (\ref{150626a}) and (\ref{050104b}) for the rotational motion (without dissipation).} $\theta$ is initially set equal to $f$ so the sections go through the middle of the synchronous resonance ($\dot{\gamma}=0$). 
\bfx{All the remaining variables are initially set to zero.}
%\bfx{The initial $\psi =0^\circ$, and for the initial velocities $\dot \vp = \dot \psi = \w_\ia = \w_\jb = 0$.}
The dots are blue for stable rotation (libration \bfx{or circulation}), and red for chaotic motion. The distinction between stable and unstable trajectories is based on the estimate of the second derivative of $\xi$  with respect to the initial condition $\dot\gamma(0)$ \citep[see][]{Laskar_1993PD}. The plateaus correspond to the resonance crossings ($\xi/\nu = k/2$). The red dots randomly distributed between the plateaus indicate the overlapping of the associated spin-orbit resonances, which generates chaotic motion.
\label{fig:section} }
\end{center}
\end{figure*}

A common way of visualizing and understanding the different regimes is to use frequency map analysis \citep{Laskar_1993PD}.
We selected two representative pairs ($ \alpha, \sigma$) and plotted the corresponding cross-section maps \bfx{for the three initial obliquities} in Figure~\ref{fig:section}. 
We are able to identify the stable and the chaotic regions very easily.
We observe that stable islands can exist in the middle of chaos, implying a large diversity of possibilities for the final rotation.
In the chaotic regions for the rotation, the obliquity is also chaotic and can vary between $0^\circ$ and $180^\circ$.
\bfx{For high obliquities, the size of the stable synchronous island shrinks and the chaotic zone is extended. 
Therefore, depending on the obliquity, the rotation alternates between more or less chaotic.
In times of small obliquity, the rotation can be trapped in a stable resonant island, preventing the obliquity from increasing again, and the spin can become stable (see Fig.\,\ref{fig:tides}).}

For small values of $\alpha$ and/or $\sigma$, the individual resonances (Eq.\,(\ref{150616z})), associated with the plateaus in Figure~\ref{fig:section}, are well separated.
Thus, for dissipative systems, the rotation can be captured in individual spin-orbit resonances and stay there.
For rotation rates increasing from lower values, the sub-synchronous resonance with $\dot\theta=n-\nu = 2n-\nb$ has the largest amplitude, so this is the most likely spin-orbit resonance to occur.
For rotation rates decreasing from higher values, the super-synchronous resonance with $\dot\theta=n+\nu=\nb$ becomes the most likely possibility.
This case is very interesting as it corresponds to a synchronous rotation with the inner orbit period.
However, since the amplitude of this resonance is small, the rotation can easily escape it and subsequently evolve into the classic synchronous rotation with the outer orbit ($\dot\theta = n$).

The amplitude of the resonant terms increases with $\alpha$ and $\sigma$.
At some point, the individual libration islands merge and chaotic motion can be expected (see section~\ref{chaos}).
The transition between the two regimes is roughly given by the dashed curved, obtained with expression (\ref{150618c}).
Near the transition regime, chaotic rotation can be observed around the separatrix of the synchronous resonance, but the motion is still regular nearer the center (Fig.~\ref{fig:section}, left).
Therefore, when the rotation of a body is evolving %(or increasing) 
by tidal effect, the rotation becomes chaotic when approaching the synchronous resonance.
However, depending on the strength and on the geometry of the tidal torque, it is still possible to find a pathway to the synchronous rotation.
The wandering in the chaotic region may also provide a path into another stable super- or sub-synchronous resonance (see Fig.\,\ref{fig:tides}).

For $\alpha > 0.4$ and $\sigma \sim 1$, there is a large overlap between several resonant terms, in particular for the negative ones ($k<0$).
As a consequence, there is a large chaotic zone for the spin (Fig.~\ref{fig:section}, bottom).
Some of the individual resonances may still subsist, including the synchronous one, but they have small stable widths and they \bfx{can only be reached at periods of small obliquity}.
\bfx{The chaotic motion is maximized for $\sigma \approx 1$, which corresponds to $(B-A)/C \approx 0.35$, because secondary resonances generate strong instabilities inside the synchronous island \citep{Robutel_etal_2012}.} %inside the synchronous island.
Finally, for $\alpha \approx 0.49$ the system reaches the 3/1 mean motion resonance, which introduces additional forcing to the rotation, and all trajectories become chaotic.
%For large $\alpha$ values, capture in stable higher order spin-orbit resonances that lay outside the chaotic zone is also possible. 

%%%%%%%%%%%%%%%%%%%%%%%%%%
\section{Application to the Pluto-Charon system}

In 1978, regular series of observations of Pluto showed that the images were consistently elongated \bfx{revealing} the presence of Charon \citep{Christy_Harrington_1978}.
The orbital parameters hence determined have shown that the two bodies evolved in an almost circular orbit with a 6.4-day period \citep[e.g.,][]{Tholen_Buie_1997}.
Charon has an important fraction of the total mass (about 11\%), and therefore the system is considered  a binary planet rather than a planet and a moon.
Later, it was found that four tiny satellites  move around the barycenter of the Pluto-Charon system \citep{Weaver_etal_2006, Brozovic_etal_2015}, also in nearly circular and coplanar orbits (see Table~\ref{tab:orbs}).

\begin{table}
\begin{center}
\begin{footnotesize}
\caption{Mean orbital and physical parameters for Pluto's moons \citep{Brozovic_etal_2015}. The parameter $m_0$ is the mass of Pluto.}
\label{tab:orbs}
\begin{tabular}{|c|c|c|c|c|c|}
\hline
 Param. & Charon & Styx & Nix & Kerberos & Hydra\\
\hline
\hline
$P$\,[day]\dotfill  & $6.3872$&$20.1617$ & $24.8548$ & $32.1679$&$38.2021$  \\
$\nb/n$\,\dotfill & $1$ & $3.1565$ &$3.8913$ &$5.0363$ & $5.9810$\\
$\alpha$\,\dotfill & $1$ & $0.46203$ &$0.40246$& $0.33932$&  $0.30278$ \\
$e$\,\dotfill &$0.00005$ &$0.00001$ &$0.00000$ & $0.00000$& $0.00554$\\
%$m/m_P$\,\dotfill &$0.12176$ & $0\,10^{-7}$&$3.4499\,10^{-6}$ &$1.264910^{-6}$ & $3.6798\,10^{-6}$\\
$\m/m_0$\,\dotfill &$0.12176$ & $0\,10^{-7}$&$3.4\,10^{-6}$ &$1.3\,10^{-6}$ & $3.7\,10^{-6}$\\
$R$\,[km]\dotfill  &$604$& $4-14$& $23-70$ & $7-22$& $29-86$   \\
\hline
\end{tabular}\\
\end{footnotesize}
\end{center}
\end{table}

%The normalised angular momentum density of the system, assuming equal densities, is 0.45 \citep{McKinnon_1989}, exceeding the critical value 0.39, above which no stable rotating single object exists \citep[e.g.,][]{Durisen_Tohline_1985}.
%The proto-planetary disk is unlikely to produce such systems.
%The prevailing theory is that Charon resulted from a giant impact of two proto-planets in the early inner Kuiper belt \citep{Stern_etal_2006, Canup_2011}.
%The small satellites likely accreted from debris of that collision that were ejected from the binary orbit \citep{Kenyon_Bromley_2014, Walsh_Levison_2015}.

Pluto's brightness also varies with a period of 6.4~days \citep[e.g.,][]{Tholen_Tedesco_1994}.
Since Charon is too dim to account for the amplitude of the variation, this period has been identified as the rotation of Pluto.
Therefore, at present, the spin of Pluto is synchronous with the orbit of Charon, a configuration acquired from the action of tidal torques raised on Pluto by Charon \citep[e.g.,][]{Farinella_etal_1979, Cheng_etal_2014a}.
Tidal torques raised on Charon and on the remaining satellites by Pluto are even stronger, so all these bodies are presumably also tidally evolved.

\citet{Showalter_Hamilton_2015} \bfx{have measured} the brightness variations of the small satellites.
They concluded that Hydra and Nix show no obvious pattern, suggesting that their rotation is chaotic.
Assuming that these two satellites are uniform triaxial ellipsoids, they additionally estimate their figure semi-axis ratios, which allows us to compute \bfx{$B/C \approx 0.92$ and $A/C \approx 0.44$ for Hydra, and $B/C \approx 0.94$ and $A/C \approx 0.31$ for Nix.}
%\be
%\left( \frac{B-A}{C} \right)_\mathrm{Hydra} = 0.49 \pm 0.20
%\ , \quad %\mathrm{and} \quad
%\left( \frac{B-A}{C} \right)_\mathrm{Nix} = 0.63 \pm 0.13 \ .
%\llabel{150617a}
%\ee
%\be
%\frac{\sigma_\mathrm{Nix}}{n} = 1.38 \pm 0.13
%\ , \quad \mathrm{and} \quad
%\frac{\sigma_\mathrm{Hydra}}{n} = 1.21 \pm 0.22  \ .
%\llabel{150617b}
%\ee

In Figure~\ref{fig:global} we already show the global dynamics for the Pluto-Charon system, for which $\delta = 0.1085 \approx 0.1$ \citep{Brozovic_etal_2015}.
Therefore, we plot vertical lines at the $\alpha$ values corresponding to the small satellites in the system (Table~\ref{tab:orbs}).
Moreover, since we have an estimation for the axial asymmetries of Hydra and Nix, %(Eq.\,(\ref{150617a})) 
we also show a dot for their $(B-A)/C$ values in Figure~\ref{fig:global}.
These two satellites are in a different dynamical regime:  the rotation of Hydra can present a stable spin-orbit coupling (Fig.\,\ref{fig:section}, top), while the rotation of Nix is most likely chaotic (Fig.\,\ref{fig:section}, bottom), as discussed in section~\ref{glodyn}.
The spin dynamics of Kerberos is probably similar to that of Hydra, while the rotation of Styx can be even more chaotic than that of Nix, depending on their $(B-A)/C$ values. 
%In Figure~\ref{fig:section}(c) we used 0.5 as example.

To test the reliability of the dynamical picture described in the previous section, we now add the effect from tides to our model, \bfx{so that we can follow the long-term evolution of the spin of these satellites.}
\bfx{For simplicity,} we adopt a \bfx{constant time-lag linear model}, whose contribution to the spin is given by expression (\ref{150626c}).
In Figure~\ref{fig:tides} we show some examples for the final evolution of \bfx{Hydra's spin} with tidal dissipation, starting with slightly different initial values of $\theta$. 
We adopt an initial retrograde rotation of 4.4~days ($\dot\gamma/\nu=-2.3$), \bfx{and $30^\circ$ for the initial obliquity}, to force the rotation to cross the large amplitude sub-synchronous resonances.
An initial retrograde rotation for such small bodies is as likely as a prograde one \citep[e.g.,][]{Dones_Tremaine_1993}.
For the uncertain parameters we used $R=45$~km, $k_2 / Q = 10^{-4}$, and $C/(mR^2) = 0.4$.

In one example (Fig.\,\ref{fig:tides}a) the rotation of \bfx{Hydra} is trapped in the spin-orbit resonance with $\dot\gamma = - 3\nu/2$, while in the others the rotation reaches the chaotic zone.
However, after some wandering in this zone some simulations are able to find a path into \bfx{a stable spin-orbit resonance with $k>0$ (Fig.\,\ref{fig:tides}b) or with $k<0$ (Fig.\,\ref{fig:tides}c).
Interestingly, while for resonances with $k>0$ the spin axis stabilizes near zero degrees, for resonances with $k<0$, the spin axis stabilizes with a high obliquity value.}
The last example (Fig.\,\ref{fig:tides}d) remained chaotic for the length of the integration and therefore can represent the observed present state.

\bfx{Over 10~Myr, most of our simulations remained chaotic, but several captures in stable non-synchronous resonances also occurred.}
The final scenario depends on the initial conditions, and also on the tidal model.
\bfx{A constant-Q model would prevent any capture in resonance \citep{Goldreich_Peale_1966}, while a visco-elastic model would increase the chances of capture \citep[e.g.,][]{Makarov_2012, Correia_etal_2014}.
Strange attractors can also exist in the chaotic zone, which may prevent the spin from stabilizing \citep[e.g.,][]{Batygin_Morbidelli_2011}}.
However, the global dynamics described in section~\ref{glodyn} is very robust and reliable, since it does not depend on dissipative forces (Fig.\,\ref{fig:global}).

%%%%%%%%%%%%%%%%%%%%%%%%%%
\section{Discussion}

%In this paper we have shown that non-synchronous spin-orbit coupling and chaotic rotation is possible for circumbinary bodies.
This work was motivated by the recent observations on the rotation of Hydra and Nix \citep[][]{Showalter_Hamilton_2015}, which appear to be chaotic.
\bfx{Our model confirms that chaotic rotation is a likely scenario for both satellites, but stable spin-orbit coupling could also be possible, in particular for Hydra.}
\bfx{This model assumes a hierarchical three-body system with coplanar orbits.
However, we also integrated the spins of Hydra and Nix using the ephemerides for the full system provided by \citet{Brozovic_etal_2015} and SPICE routines \citep{Acton_1996}, and we have found no evidence of any substantial differences.}
%The age of the Pluto-Charon system is unknown, but assuming that is similar to that of the Earth, \bfx{Hydra} could have evolved into a stable spin-orbit resonance as well.
%One simple possibility is that an impact occurred in a recent past, temporarily broke the synchronous equilibrium, as for the planet Mercury \citep[][]{Correia_Laskar_2012}.

In our study, we conclude that stable spin-orbit coupling is also a plausible scenario for near circular circumbinary bodies with small $\alpha$ and/or small $\sigma$ (Fig.\,\ref{fig:global}).
Equilibrium rotation occurs for $\dot\theta=n + k\nu/2$, with $\nu = \nb-n$ and $k\in\mathbb Z$.
The largest amplitude non-synchronous resonance corresponds to the sub-synchronous resonance at $n-\nu=2n-\nb$.
Bodies captured in this resonance %either present a rotation period longer than the orbital one, or 
present retrograde rotation if $\nb/n > 2$.
This condition is verified for \bfx{the small satellites of the Pluto-Charon system (Table~\ref{tab:orbs}), and likely for any circumbinary system, since large orbital instabilities are expected for period ratios below the 2/1 mean motion resonance}.
Therefore, bodies trapped in a non-synchronous resonance are also likely to present retrograde rotation.

Lately, many planets have been detected around binary stars \citep[e.g.,][]{Welsh_etal_2012}.
So far, all these planets are gaseous giants, for which the axial asymmetry is very low, for instance $(B-A)/C \sim 10^{-7}$ for Jupiter \citep{Jacobson_2001}.
Therefore, although many of these planets are close enough to their stars to undergo tidal dissipation, spin-orbit coupling is very unlikely.
However, for smaller mass Earth-like circumbinary planets these states are possible since $(B-A)/C \sim 10^{-5}$ \citep[e.g.,][]{Yoder_1995cnt}. 
In particular, non-synchronous rotation is possible, which is an important point to take into account in future habitability studies \citep[e.g.,][]{Selsis_etal_2007}.
Unlike the Pluto-Charon system, circumbinary exoplanets usually present eccentric orbits.
As a consequence, the number of spin-orbit resonances drastically increases (Eq.\,(\ref{150616w})).
Overlap of the different contributions is then easier, so chaotic rotation can also be more likely in this case.

%%%%%%%%%%%%%%%%%%%%%%%%%%
\begin{acknowledgements}
We thank Doug Hamilton and Mark Showalter for helpful suggestions.
We acknowledge support from the ``conseil scientifique'' of the Observatory of Paris and CIDMA strategic project UID/MAT/04106/2013.
\end{acknowledgements}

\bibliographystyle{aa}
\bibliography{correia}

\begin{figure*}
%%%%%%%%%%%%%%%%%%%%%%%%%%%%%%%%%%%%%%%%%%%%%
\begin{center}
\includegraphics[width=.9\textwidth]{\figpath 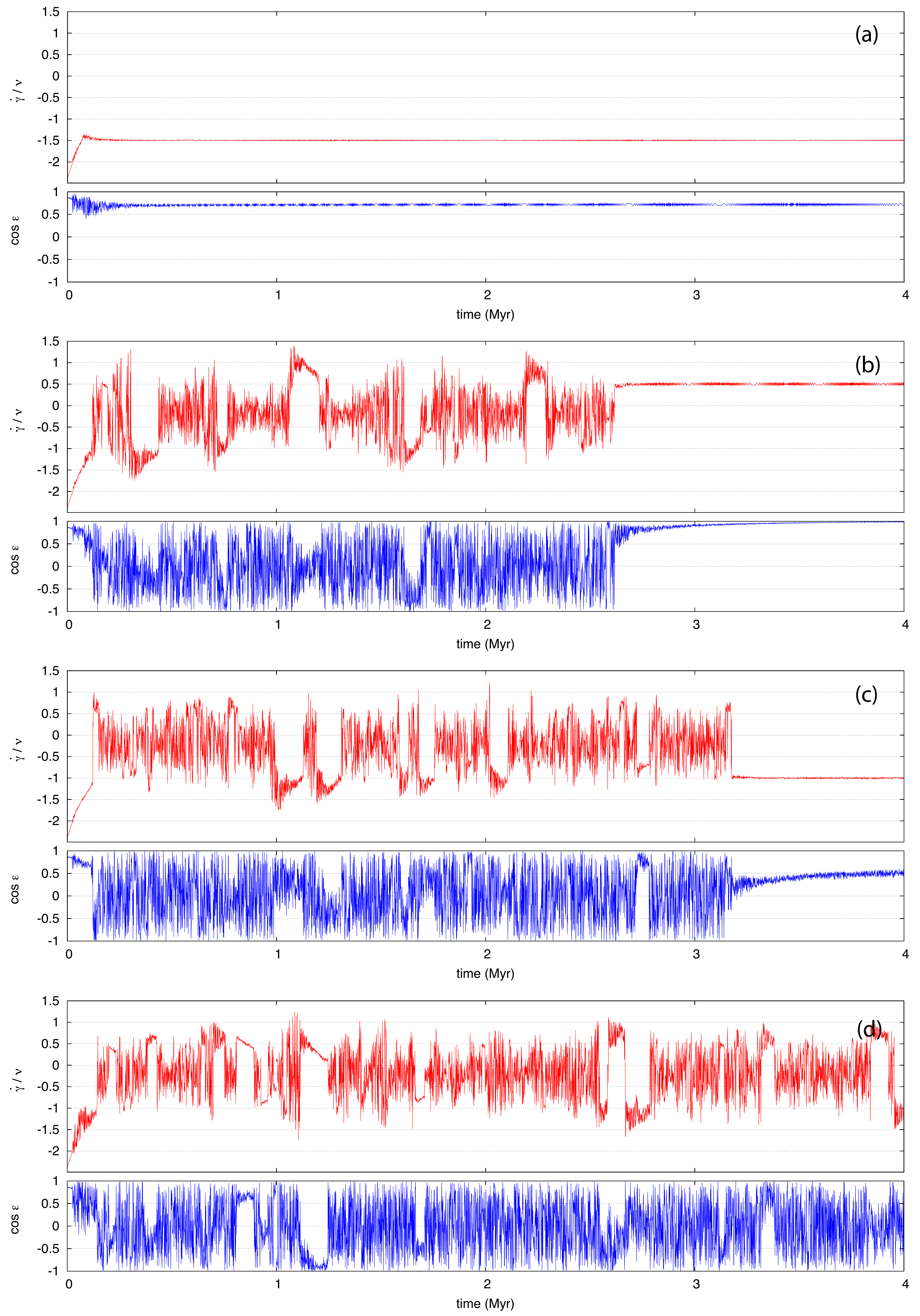} 
  \caption{Possible evolutions for the \bfx{spin of Hydra.
  We show the rotation $\dot\gamma / \nu$ (in red) and the cosine of the obliquity (in blue).}
  We numerically integrated the equations of the three-body problem together with equations (\ref{150626a}), (\ref{050104b}), and (\ref{150626c}) for the rotational motion.
We adopt an initial retrograde rotation of 4.4~days with different initial values of $\theta$, and an initial obliquity of $30^\circ$.
%Dissipation is set to a high value so that we can speed-up the simulations. 
   \llabel{fig:tides}   }
\end{center}
\end{figure*}

\appendix

\newpage
%%%%%%%%%%%%%%%%%%%%%%%%%%
\section{Equations of motion}
\llabel{eom}

We let ($\vi, \vj, \vk$) be a non-inertial frame attached to the body's principal inertial axes, with inertia tensor $\vv{I} = diag(A, B, C)$.
In this frame, $\vw = (\w_\ia, \w_\jb, \w_\kc)$ and the angular momentum $\vL =  (A \w_\ia, B \w_\jb, C \w_\kc)$.
The equations of motion are given by
\be
\vv{I} \cdot \dot{\vw} + \vw \times \vL = \vv{T}_0 +  \vv{T}_1 \ ,
\llabel{150626a}
\ee
where $\vv{T}_i$ is the gravitational torque on the body's figure.
For second order interactions, we have \citep[e.g.,][]{Goldstein_1950}
\be
\vv{T}_i = \frac{3 G m_i}{r_i^5} \left[ (B-A) \, y_i \,
\vr_i \times \vj + (C-A) \, z_i \, \vr_i \times \vk \right] \ ,
\llabel{050104b}
\ee
where $\vr_i$ has coordinates $(x_i, y_i, z_i)$ in the body's frame.
Thus, projecting equation (\ref{150626a}) over each axis ($\vi, \vj, \vk$) gives
\be
\dot \omega_\ia = \frac{C-B}{A} \left( \frac{3 G m_0}{r_0^5}  \, y_0 \, z_0 + \frac{3 G m_1}{r_1^5}  \, y_1 \, z_1 - \w_\jb \, \w_\kc \right) \ ,
\llabel{150703a}
\ee
\be
\dot \omega_\jb = \frac{A-C}{B} \left( \frac{3 G m_0}{r_0^5}  \, z_0 \, x_0 + \frac{3 G m_1}{r_1^5}  \, z_1 \, x_1 - \w_\kc \, \w_\ia \right) \ ,
\llabel{150703b}
\ee
\be
\dot \omega_\kc = \frac{B-A}{C} \left( \frac{3 G m_0}{r_0^5}  \, x_0 \, y_0 + \frac{3 G m_1}{r_1^5}  \, x_1 \, y_1 - \w_\ia \, \w_\jb \right) \ .
\llabel{150626b}
\ee

To solve these equations, a set of generalized coordinates to specify the orientation of the satellite must be chosen. 
We adopt the modified Euler angles ($\theta, \vp, \psi$) as defined in \citet{Wisdom_etal_1984}. 
Starting the $\vk$ axis coincident with the normal to the orbital plane, and $\vi$ along 
%the direction of the pericentre, 
an inertial direction,
we first rotate the body about the $\vk$ axis by an angle $\theta$, then we rotate about $\vi$ by an angle $\vp$, and finally we rotate about the $\vj$ by an angle $\psi$.
Then, 
%we can express
\be
   x_i / r_i  = \cos(\theta-f_i) \cos\psi - \sin(\theta-f_i) \sin\psi \sin\vp \ ,
 \llabel{150703c}
\ee 
\be
   y_i / r_i  = - \sin(\theta-f_i) \cos\vp \ , 
 \llabel{150703d}
\ee 
\be
   z_i / r_i  = \cos(\theta-f_i) \sin\psi + \sin(\theta-f_i) \cos\psi \sin\vp \ , 
\llabel{150626x}
\ee
and
%\be
%   \w_\ia = - \dot\theta \sin\psi \cos\vp + \dot\vp \cos\psi  \ ,
% \llabel{150708a}
%\ee 
%\be
%   \w_\jb = \dot\theta \sin\vp + \dot\psi  \ , 
% \llabel{150708b}
%\ee 
%\be
%   \w_\kc = \dot\theta \cos\psi \cos\vp + \dot\vp \sin\psi   \ .
% \llabel{150625w}
%\ee 
\be
   \dot\theta = (\omega_\kc \cos\psi - \omega_\ia \sin\psi) / \cos\vp  \ ,
\llabel{150627b}
\ee 
\be
   \dot\vp = \omega_\kc \sin\psi + \omega_\ia \cos\psi \ , 
\llabel{150703e}
\ee 
\be
   \dot\psi = \omega_\jb - (\omega_\kc \cos\psi - \omega_\ia \sin\psi) \tan\vp \ . 
\llabel{150703f}
\ee 

The full equations of motion for the spin are thus described by the set of variables ($\w_\ia, \w_\jb, \w_\kc, \theta, \vp, \psi$), whose derivatives are given by equations (\ref{150703a})$-$(\ref{150626b}) together with (\ref{150627b})$-$(\ref{150703f}).
The angle $\ve$ between the $\vk$ axis and the normal to the orbit, usually called obliquity, can be obtained from 
\be
\cos \ve = \cos \vp \cos \psi \ .
 \llabel{150715a}
\ee

For the dissipative tidal torque, we adopt a constant time-lag linear model, whose contribution
to the spin (Eq.\,(\ref{150626a})) is given by \citep[e.g.,][]{Mignard_1979}
\be
\vv{T}_d =  3 k_2 G R^5 \Delta t \sum_{i=0,1} \frac{m_i^2}{r_i^8} \left[ ( \vr_i \cdot \vw ) \, \vr_i - r_i^2 \vw + \vr_i \times \dot{\vr_i}  \right] \ , 
\llabel{150626c}
\ee
where $k_2$ is the Love number, $\Delta t$ is the time lag, $Q^{-1}\equiv n\Delta t$ is the dissipation factor, and $R$ is the radius of the rotating body.
Also note that $\dot{\vr_i}$ is the derivative of $\vr_i$ in an inertial reference frame.

%Projecting on each axis, we get the additional contributions for the angular accelerations
%\be
%\dot \omega_\ia = \frac{K}{A} \sum_{i=0,1} \frac{m_i^2}{r_i^8} \left[ ( \vr_i \cdot \vw ) \, x_i - r_i^2 \w_\ia + (y_i \, \dot z_i - z_i \, \dot y_i ) \right]   \ .
%\llabel{150703g}
%\ee
%\be
%\dot \omega_\jb = \frac{K}{B} \sum_{i=0,1} \frac{m_i^2}{r_i^8} \left[ ( \vr_i \cdot \vw ) \, y_i - r_i^2 \w_\jb + (z_i \, \dot x_i - x_i \, \dot z_i ) \right]  \ ,
%\llabel{150703h}
%\ee
%\be
%\dot \omega_\kc  = \frac{K}{C} \sum_{i=0,1} \frac{m_i^2}{r_i^8} \left[ ( \vr_i \cdot \vw ) \, z_i - r_i^2 \w_\kc + (x_i \, \dot y_i - y_i \, \dot x_i ) \right]  \ .
%\llabel{150626e}
%\ee

\section{Series with eccentricity}
\llabel{swe}

Equation (\ref{150616a}) for the rotational motion can be expanded in power series of the eccentricities and semi-major axis ratio $\alpha$ as given by the general expression (\ref{eq:rot_gen})
\be
\ddot \theta = -\frac{\sigma^2}{2} \Gamma \ ,
\llabel{150616az}
\ee
with
\be
\Gamma = \sum_{l,p,q} \beta_{l,p,q} (e, \eb, \delta, \alpha) \sin2 (\theta - p M - q \Mb) \ .
\llabel{150621a}
\ee
When we truncate the series to the first order in the eccentricities and to the second order in $\alpha$, % i.e., neglecting terms in ${\cal O} (e^2, e_b^2, \alpha^3)$ 
we get

\begin{eqnarray}
\Gamma  &=&  \big( 1 + \frac54\vo_2  \big) \sin (2 \theta - 2 M) \nonumber \\
&& + \frac38\vo_2\sin (2 \theta - 2 \Mb )  \nonumber \\
&& +\frac{35}{8}\vo_2\sin (2\theta - 4 M + 2 \Mb) \nonumber \\
&& + \big(\frac72   + \frac{45}{8} \vo_2 \big) \, e \sin( 2\theta -3M)  \nonumber \\
&& - \big(\frac12   - \frac{5}{8} \vo_2 \big) \, e \sin( 2\theta -M)  \nonumber \\
&& -\frac{105}{16}  \vo_2 \, e \sin(2\theta - 3M + 2\Mb)  \nonumber \\
&& +\frac{455}{16} \vo_2 \, e \sin(2\theta - 5M + 2\Mb)  \nonumber \\
&& +\frac{15}{16} \vo_2 \, e \sin(2\theta + M- 2\Mb)  \nonumber \\
&& +\frac{15}{16} \vo_2 \, e \sin(2\theta - M- 2\Mb)  \nonumber \\
&& +\frac{3}{8} \vo_2 \, \eb \sin(2\theta - 3\Mb) \\
&& -\frac{9}{8} \vo_2 \, \eb \sin(2\theta - \Mb)  \nonumber \\
&& -\frac{105}{8} \vo_2 \, \eb \sin(2\theta - 4M + \Mb)  \nonumber \\
&& +\frac{35}{8} \vo_2 \, \eb \sin(2\theta - 4M + 3\Mb)  \nonumber \\
&& -\frac{5}{4} \vo_2 \, \eb \sin(2\theta - 2M -\Mb)  \nonumber \\
&& -\frac{5}{4} \vo_2 \, \eb \sin(2\theta - 2M +\Mb)  \nonumber  \ .
\llabel{150616w}
\end{eqnarray}

\end{document}